\documentclass[aps,prc,twocolumn,fleqn]{revtex4}

\usepackage{amsmath,amssymb,amsopn}
\usepackage{epsfig}

\DeclareMathOperator{\tr}{\mathrm{Tr}}
\newcommand{\tauth}{\tau_\mathrm{th}}
\newcommand{\taudec}{\tau_\mathrm{dec}}

\begin{document}

\title{Decoherence and Entropy Production in Relativistic Nuclear Collisions}

\author{Rainer J. Fries}
\affiliation{Cyclotron Institute and Department of Physics, 
Texas A{\&}M University, College Station, TX 77801}
\affiliation{RIKEN/BNL Research Center, Brookhaven National Laboratory, 
Upton NY 11973}
\author{Berndt M\"uller}
\affiliation{Department of Physics, Duke University, Durham, NC 27708}
\author{Andreas Sch\"afer}
\affiliation{Institut f\"ur Theoretische Physik, Universit\"at Regensburg, 
93040 Regensburg, Germany}

\date{\today}

\begin{abstract}
Short thermalization times of less than 1 fm/$c$ for quark and gluon matter
have been suggested by recent experiments at the Relativistic Heavy
Ion Collider (RHIC). It has been difficult to justify this rapid 
thermalization in first-principle calculations based on perturbation theory
or the color glass condensate picture. Here, we address the related 
question of the decoherence of the gluon field, which is a necessary
component of thermalization. We present a simplified  leading-order computation of 
the decoherence time of a gluon ensemble subject to an incoming flux
of Weizs\"acker-Williams gluons. We also discuss the entropy produced 
during the decoherence process and its relation to the entropy in the 
final state which has been measured experimentally.
\end{abstract}

\maketitle

\section{Introduction}

Collisions of nuclei at very high energies have been studied at
the Relativistic Heavy Ion Collider (RHIC) in recent years, in order
to explore the formation and the properties of the quark gluon plasma 
(QGP). One striking discovery was the fact that ideal hydrodynamics 
could describe many salient features of the expansion and cooling of the 
fireball \cite{whitepaper:05,hydro,Kolb:2003dz}. In particular the 
azimuthal asymmetry in collisions with nonzero impact parameter requires 
an onset of the hydrodynamic expansion at rather early times, between 0.5 
and 1 fm/$c$. This represents a puzzle since the application of ideal 
hydrodynamics mandates complete thermalization of the system.

Scenarios based on the picture of a dilute system of perturbatively interacting 
partons and minijets fail to describe early thermalization \cite{Baier:2000sb}
and the formation of anisotropic transverse flow \cite{Molnar:2001ux}. 
Multi-parton interactions, either involving ternary collisions 
\cite{Wong:1996va,Xu:2007aa} or collective effects mediated by plasma
instabilities \cite{Mrowczynski:2005ki} have been proposed as solutions
to this problem. However, both mechanisms require a large initial entropy
production before they can become effective. The color glass condensate 
model \cite{mv,cgc} seemingly offers a viable explanation of the conundrum. 
It introduces the saturation scale $Q_s$ which sets the scale for all dynamical
processes. For RHIC \cite{Krasnitz:2003jw} the magnitude of $Q_s^2$ is 
estimated to be approximately 2 GeV$^2$, and thus thermalization times of 
order $1/Q_s$ do not seem impossible. However, the production of entropy
is a nontrivial problem in any model that is based on the assumption of
the dominance of classical fields in the initial state, like the color glass
condensate model \cite{Kharzeev:2006zm,Romatschke:2006nk}. 

The situation is further complicated by the fact that it is not clear how 
close quark-gluon matter must be to complete thermalization for 
hydrodynamics to be successful. It is also conceivable that a hydrodynamic 
evolution starting at a later time and supplemented by 
other mechanisms of transverse dynamics during the long
{\it off-equilibrium} phase, can give an equally good or even better 
description of the data. For example, the anisotropic collective transverse 
flow may be generated, in part, by interactions of minijets with the bulk
medium \cite{Hwa:2007ae} or by anisotropies in the initial gluon field
\cite{Fries:2007iy}. Hydrodynamic calculations with viscous 
corrections describing small deviations from equilibrium start to 
become available \cite{Baier:2006gy,Dusling:2007gi,Song:2007ux}
and may soon help to test this possibility quantitatively.

One necessary ingredient for thermalization is the decoherence of the
initial gluon field. Coherent fields can lead to large anisotropies in
pressure and even negative pressure, which are symptoms of a 
state very far from thermal equilibrium. Thus the decoherence time 
$\taudec$ should be even smaller than the equilibration time $\tauth$.
We argue that 
the fundamental process at work is somewhat analogous
to Coulomb explosion imaging, see 
\cite{deco1}, used routinely in molecular physics.
If a molecule transverses a very thin metal foil all bonds are
broken, the ions decohere and fly apart. From the momentum
distribution of the fragments one can then extract information
on the original wavefunction. In heavy ion collisions each Lorentz
contracted ion acts like such a foil for the other.

That the loss of information due to decoherence can generate a rapid increase
in entropy in early phases of heavy ion collisions was realized already
early on, e.g. by Elze \cite{deco2}. The fact that the hydrodynamic
evolution is known to be very close to the ideal one and thus isentropic
in heavy ion collisions, further stresses the need for massive entropy
production in very early phases. Indeed, two of us have argued
in \cite{deco3}, that decoherence can easily generate a large fraction of
the total produced entropy. However, the question of the appropriate
time scale of entropy production remained open.

A computation of the decoherence time in leading order in perturbation
theory was recently presented by two of us in \cite{Muller:2005yu}. 
Here, we want to strengthen this argument by presenting a
calculation of the decoherence time as a function of the gluon 2-point 
function in the nucleus. We then proceed to evaluate our general result 
within the framework of the McLerran-Venugopalan model \cite{mv}. 
This is made possible by new results for the effects of the running
of the coupling constant on the color glass condensate \cite{Kovchegov:2007vf},
which solve a hitherto unresolved UV problem. Our result agrees with that of 
Ref.~\cite{Muller:2005yu} within theoretical uncertainties and suggests that, 
indeed, $\tau_{\rm dec} \sim Q_s^{-1}$, and that $\tau_{\rm dec}$ is
numerically smaller than 1 fm/$c$ at RHIC. In Sec.\ V we can then revisit 
some of our previous arguments \cite{deco3} about entropy production in 
the framework of the short decoherence times at RHIC.

\section{The Decoherence Time}

We describe the gluons in a nucleus by a density operator $D$. This nucleus 
is subjected to the incident  gluon field of a second, large and very fast 
nucleus scattering off it. We compute the time evolution of the density operator 
$D$ under the influence of the perturbation presented by the second nucleus. 
In the following we denote the initial unperturbed gluon field of the first
nucleus with $A'$, the final gluon field with $A$ and the field of the second
nucleus with $B$.
The decoherence time of the gluon field $A$ is defined as the inverse decay 
time of the ratio 
\begin{equation}
  \frac{\tr D^2(t)}{[\tr D(t)]^2}.
\end{equation}

Let us introduce some useful notations.
We deal with matrix elements $D_{\hat A,A} = \langle \hat A|D|A\rangle$ of the 
density matrix. We can treat the final gluon field as almost on-shell for long 
times after the collision. In practice that means that we can decompose it in 
free modes $|A\rangle= |k,\lambda,a\rangle$ characterized by momentum $k$, 
polarization $\lambda$ and color $a$. We can use the same technique for the
field $B$ of the fast moving nucleus 2 which is Weizs\"acker-Williams-like.
On the other hand, the initial-state gluons $A'$ in nucleus 1 are in a bound 
state and generally off-shell. We do not attempt to describe this field in
detail. It turns out that the only two ingredients needed are an ansatz for
the matrix elements $D_{\hat A',A'}$ of the density matrix of the bound fields, 
and the matrix elements $H_{A',A}$ of the Hamiltonian coupling the fields 
$A$, $B$ and $A'$. 

\begin{figure}[tb]   
\resizebox{1\linewidth}{!}
          {\rotatebox{0}{\includegraphics{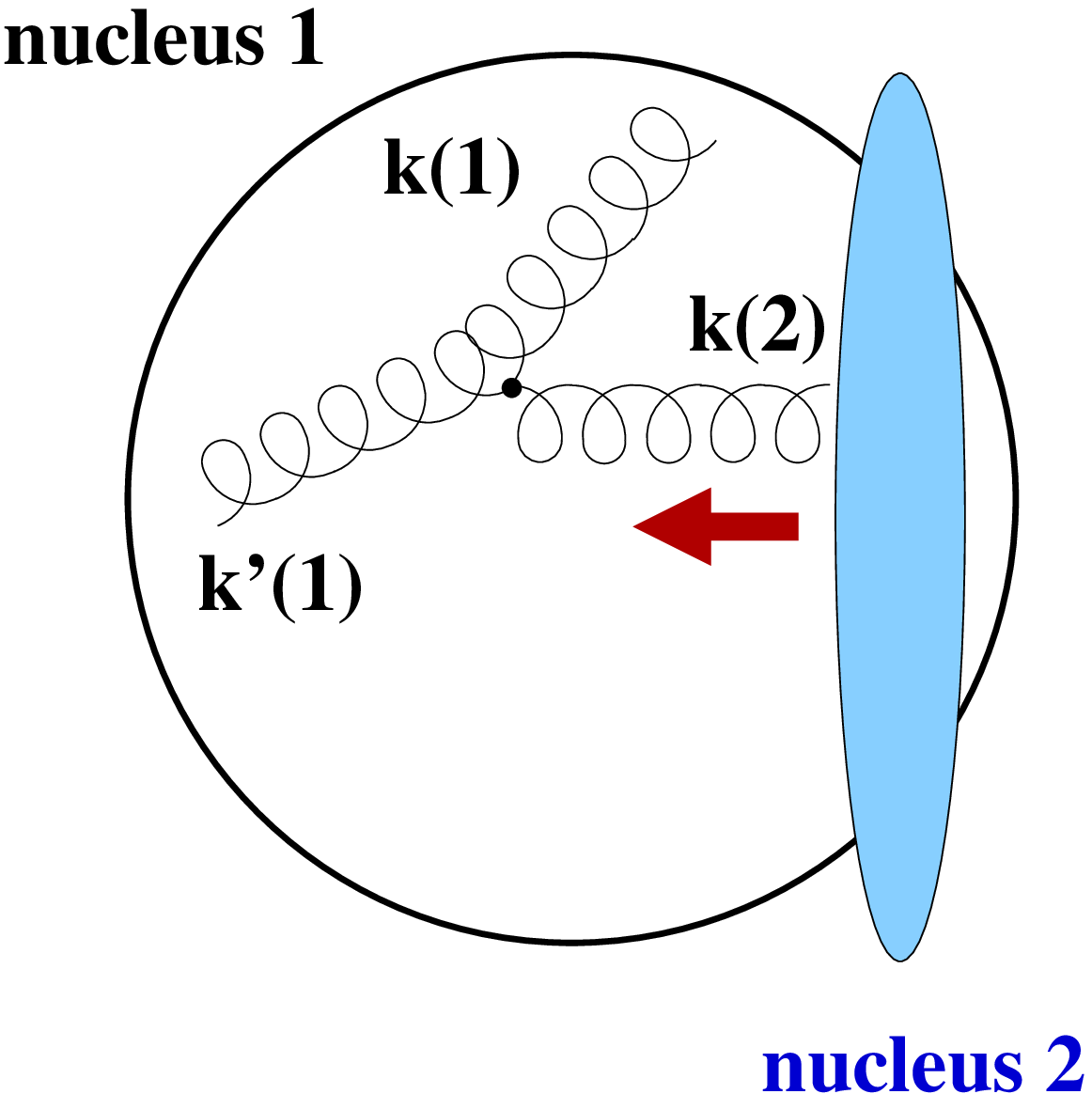}}\hskip 1 cm
          {\includegraphics{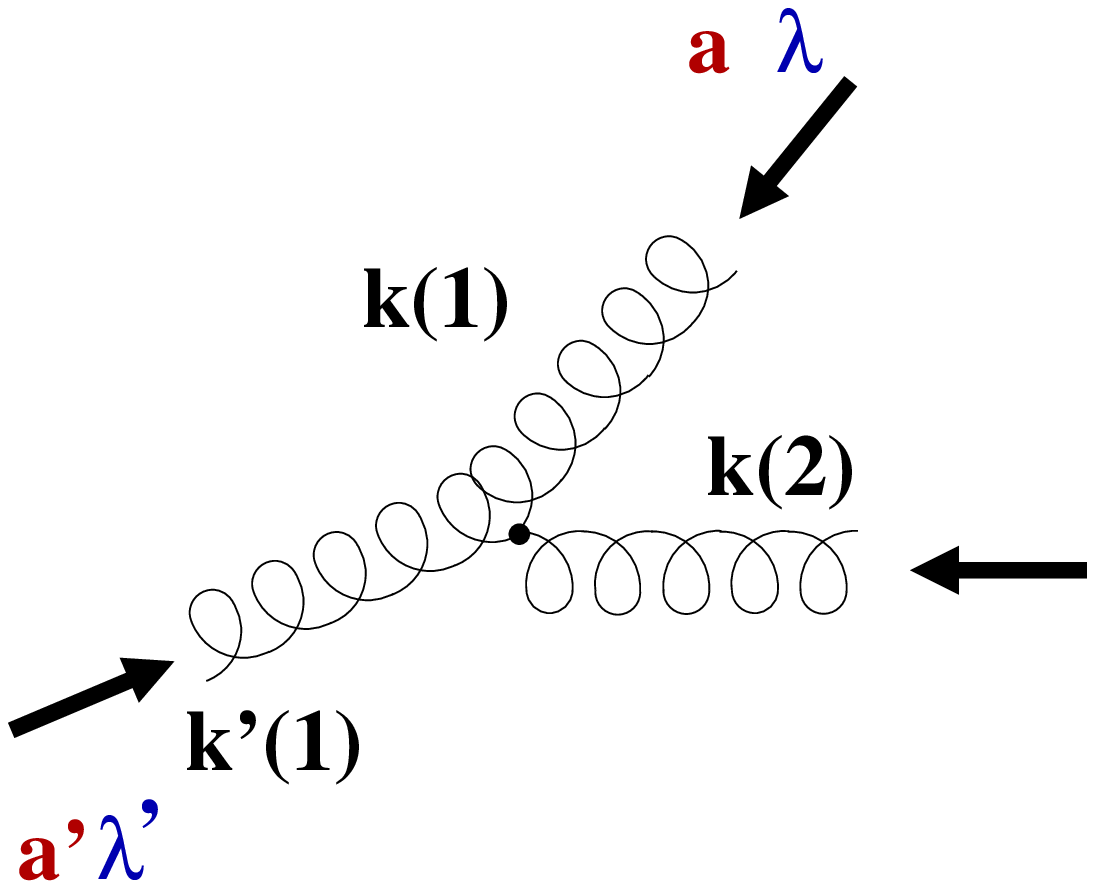}}}
\caption{Lowest order perturbative process that contributes to the 
time evolution of the matrix element $D_{\hat A,A}$ for fields $\hat A$,
$A$ which are separated from the initial fields $\hat A'$ and $A'$ by 
a rapidity gap.}
\label{fig:1}
\end{figure}

\begin{figure}[tb]   
\includegraphics[width=0.95\linewidth]{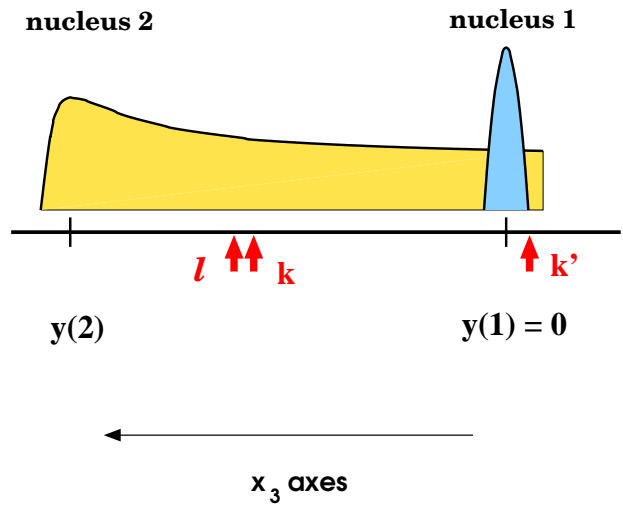}
\caption{Sketch of the relevant rapidity distributions. The rapidity
of the gluon 1 in the nucleus at rest $k'(1)$ is close to zero. The 
Weizs\"acker-Williams gluon with momentum $l$ has a nearly boost-invariant  
rapidity distribution. The distribution of the final gluon momentum $k(1)$ 
is, therefore, also nearly boost-invariant.}
\label{fig:2}
\end{figure}

We are interested in processes in which gluon modes $A'$ of the nucleus at rest 
(which are centered around rapidity $Y=0$) are scattered into modes $A$ with 
large longitudinal momenta ($Y > 1$) so that the overlap with the initial state 
is very small. In that case the leading contribution in the 
time evolution comes from second-order perturbation theory
\begin{multline}
  D_{\hat A,A} (t) = \sum_{\hat A',A'} \int_0^t d\hat t dt' \\
  \times H_{\hat A,\hat A'}(\hat t) D_{\hat A',A'}(0)H_{A',A}(t')
\end{multline}
Note that we have suppressed the field $B$ in the notation of the matrix 
elements. We treat the field $B$ rather as an external parameter given by 
the second nucleus.
The interpretation of this process is illustrated in Fig.~\ref{fig:1}. The 
rapidity distributions of the initial and final gluons are schematically shown 
in Fig.~\ref{fig:2}.

We have to specify the relevant part of the Hamiltonian matrix element $H_{A',A}$
which is the three-gluon vertex with fields $A$, $A'$ and $B$. It is given by
\begin{multline}
  H_{A',A} = i g \sum_{a',c,\lambda'} \int d^3 x \int \frac{d^4 l}{(2\pi)^4} \\
  \frac{d^4 k'}{(2\pi)^4} \frac{e^{i(k-k'-l)\cdot x}}{\sqrt{2k^+V}} 
   \times f^{ca'a} \mathcal{B}^c_\mu(l) \\  \times
   \left[ (k+k')^\mu \epsilon^{\nu *}(k,\lambda)\epsilon_\nu(k',\lambda') 
   \mathcal{A'}^{a'}(k',\lambda') \right. \\
   \qquad - (k+l)^\nu \epsilon^{\mu *}(k,\lambda) \epsilon_\nu(k',\lambda')
   \mathcal{A'}^{a'}(k',\lambda') \\
   + \left. (l-k')^\nu \epsilon_\nu^*(k,\lambda)\epsilon^\mu(k',\lambda') 
   \mathcal{A'}^{a'\mu}(k') \right] .
\label{eq:vertex}
\end{multline}
Here $t^a$ are the adjoint $SU(3)$ generators, $f^{abc}$ are the structure 
constants of $SU(3)$ and $g$ is the coupling constant. Note that we have
used Fourier transformations of the operators of the initial gluon field 
and the field of the second nucleus
\begin{align}
 A'_\nu(x) &= \sum_{a',\lambda'} \int\frac{d^4 k'}{(2\pi)^4} \left( 
 e^{-ik'\cdot x} A'^{a'}_{k',\lambda'} \right. \\ & \left. \qquad \times 
 \epsilon_\nu(k',\lambda') t^{a'} + {\rm h.c.} \right) \nonumber \\
 B^\mu (x) &= \sum_c \int\frac{d^4 l}{(2\pi)^4} \left( e^{-il\cdot x} 
 B^{c\mu}(l) t^c + {\rm h.c.} \right)
\label{eq:ft2}
\end{align}
with operators $A'^{a'}_{k',\lambda'}$ and $B^{c\mu}(l)$. This is similar 
to the usual expansion of free fields, which we apply in our case to the final
gluon field $A$.
Eq.\ \ref{eq:vertex} is then analogous to the familiar case of three 
interacting fields that are asymptotically free. However, in order to respect
the unknown dynamics of the initial field we do not specify the result of 
the operators $A'^{a'}_{k',\lambda'}$ and $B^{c\mu}(l)$ acting on the 
initial states. Rather, we use the matrix elements
\begin{align}
  \mathcal{A'}^{a'}(k',\lambda') &= \langle 0 | A^{a'}_{k',\lambda'} | A'
  \rangle  \, , \\
  \mathcal{B}^{c\mu}(l) &= \langle 0 | B^{c\mu}(l) | B \rangle \, .
\end{align}
for given initial states $|A'\rangle$ and $|B\rangle$ in Eq.\ \ref{eq:vertex}.

We now consider measurements made at times much larger than 
the time it takes for the field $A'$ to interact with the Lorentz-contracted
fast nucleus, i.e. $\int_0^t dt' \to \int_{-\infty}^\infty dt'$. 
In this case the full integral $d^4 x$ over each three gluon vertex can 
be easily carried out. 
We define the evolution matrix for the time evolution of 
the density matrix elements as
\begin{widetext}
\begin{multline}
  \label{eq:W}
  W_{AA',\hat A\hat A'} := \left\langle \int dt H_{A',A} \int dt 
  H_{\hat A',\hat A}^\dagger \right\rangle_2 \\
  =  g^2 f^{ca'a} f^{\hat c\hat a'\hat a} \frac{d^4 k'}{(2\pi)^4}\frac{d^4
    \hat k'}{(2\pi)^4} \mathcal{A'}^{a'}(k',\lambda') {\mathcal{A'}^{\hat
      a'}}^\dagger(\hat k',\hat\lambda') 
     \frac{\langle \mathcal{B}^c_\mu(k-k') {\mathcal{B}^{\hat c}_{\hat\mu}}^\dagger(\hat k - \hat k') 
     \rangle_2}{2V\sqrt{k^+\hat k^+} (VT)^2} \epsilon_\sigma^*(k,\lambda) 
   {\epsilon_{\hat\sigma}}(\hat k, \hat\lambda)\epsilon_\nu(k',\lambda')
   \epsilon_{\hat\nu}^*(\hat k',\hat\lambda')   \\
     \times  \left[ (k+k')^\mu g^{\nu\sigma}
     - (2k-k')^\nu  g^{\sigma\mu} +  (k-2k')^\sigma g^{\mu\nu} \right] 
   \left[(\hat k+\hat k')^{\hat\mu} g^{\hat\nu\hat\sigma} 
   - (2\hat k-\hat k')^{\hat\nu} g^{\hat\sigma\hat\mu}
   + (\hat k-2\hat k')^{\hat \sigma} g^{\hat\mu\hat\nu} \right] 
    \, .
\end{multline}
\end{widetext}
Since the second nucleus is moving extremely fast and we are not interested the 
time evolution of its gluon fields $B$, we have averaged over the fields in the second 
nucleus 
$\mathcal{B}^c_\mu(l) {\mathcal{B}^{\hat c}_{\hat \mu}}^\dagger 
(\hat l) \to {\langle \mathcal{B}^c_\mu (l)
{\mathcal{B}^{\hat c}_{\hat \mu}}^\dagger (\hat l) \rangle}_2$,
leaving us with an expression that only depends on the initial and final
fields in nucleus 1.

Arguing with translational and rotational invariance in the transverse
plane \cite{Lappi:06} and using that it is moving along the light cone, 
we can decompose the two-point correlation function 
of the gluon field of nucleus 2 as
\begin{multline}
  \langle \mathcal{B}_\mu^c (p) {\mathcal{B}^{\hat c}_{\hat \mu}}^\dagger (q) 
  \rangle_2 
= {\langle \mathcal{B}_\mu^c (\mathbf{p}_\perp) 
  {\mathcal{B}^{\hat c}_{\hat \mu}}^\dagger 
  (\mathbf{q}_\perp) \rangle}_2
  \frac{\pi^2\delta(p^-)\delta(q^-)}{p^+q^+} \\ 
= \delta^{c\hat c} \delta_{\mu i} \delta_{\hat \mu j} (2\pi)^2 
  \delta^2(\mathbf{p}_\perp - \mathbf{q}_\perp) \\ \times
   \frac{\pi^2\delta(p^-)\delta(q^-)}{p^+q^+} 
   \frac{p_i p_j}{p_\perp^2} G(p_\perp).
\end{multline}
where $G(p_\perp)$ is the scalar correlation function for the gluon field 
in the fast moving nucleus, and $i,j$ denote the transverse directions.

Thus far we have not made use of the fact that nucleus 1 is at rest and 
the results are valid in general as long as the phase space of initial 
and final gluons is sufficiently different. Now we note that the final
gluon momenta have large $+$ components, much larger than the original 
ones, i.e. $k^+, \hat k^+ \gg k'^+,\hat k'^+$ with
$k^{\pm}= (k^0\pm k^3)/\sqrt{2}$. Therefore
the dominant terms in Eq.\ (\ref{eq:W}) are those with the maximum 
number of factors $k^+$ or $\hat k^+$. These are
\begin{eqnarray}
  (2k-k')^\nu \epsilon_\nu(k',\lambda') &\approx & 2k^+ \epsilon^-(k',\lambda') \, , \\
  (2\hat k-\hat k')^{\hat \nu} \epsilon_{\nu*}(\hat k',\hat \lambda') 
  &\approx & 2\hat k^+ \epsilon^-_{*}(\hat k',\hat \lambda') \, .
\end{eqnarray}
Hence, the leading order contribution to the evolution operator for the gluon 
density matrix in our specific kinematic situation is
\begin{multline}
  W_{AA',\hat A\hat A'} = g^2 f^{ca'a} f^{c\hat a'\hat a} \frac{d^4
    k'}{(2\pi)^4} \frac{d^4\hat k'}{(2\pi)^4} \\ \times
  \delta(k^--k'^-)\delta(\hat k^--\hat k'^-)  
  (2\pi)^2 \delta^2 (\mathbf{k}_\perp-\mathbf{k}'_\perp -
  \mathbf{\hat k}_\perp + \mathbf{\hat k'}_\perp ) \\ \times
  G(|\mathbf{k}_\perp -\mathbf{k}'_\perp|) 
  \frac{\mathcal{P}}{2V\sqrt{k^+\hat k^+}}
  \mathcal{A'}^{a'}(k',\lambda') {\mathcal{A'}^{\hat a'}}^\dagger(\hat k',
  \hat\lambda') \\ 
  \times \epsilon^-(k',\lambda') \epsilon^{-*}(\hat k',\hat\lambda')\, ,
\end{multline}
where we have introduced the abbreviation 
\begin{equation}
  \mathcal{P} = \epsilon^i (k,\lambda) {\epsilon^j}^* (\hat k,\hat\lambda) 
  \frac{(\mathbf{k}_\perp - \mathbf{k}'_\perp)^i (\mathbf{k}_\perp - 
  \mathbf{k}'_\perp)^j}{(\mathbf{k}_\perp - \mathbf{k}'_\perp)^2}.
\end{equation}
Note that $\mathcal{P}$ is essentially the product of the projections 
of the two polarization vectors onto the transverse direction
given by the vector $\mathbf{k}_\perp-\mathbf{k}'_\perp =
\mathbf{\hat k}_\perp - \mathbf{\hat k'}_\perp$.

\section{The Gluon Density Matrix}

The evolution of the gluon density matrix of nucleus 1 can now be computed 
through
\begin{equation}
  \label{eq:dfinal}
D_{\hat A,A}(t) =  \sum_{A',\hat A'} D_{\hat A',A'}(0) 
W_{AA',\hat A \hat A'}(t) \, .
\end{equation}
For our calculation, we do not need the nuclear density matrix itself, but just the expectation
value of $ {A'}^{a'}_{k',\lambda'} {{A'}^{\hat a'}_{\hat k',\hat\lambda'}}^\dagger$
in the ground state of nucleus 1. As in \cite{Muller:2005yu} we use an 
{\em ansatz} for this expectation value which is diagonal and exhibits 
a Gaussian momentum distribution with width $1/\zeta$:
\begin{multline}
  \left\langle {A'}^{a'}_{k',\lambda'} {{A'}^{\hat a'}_{\hat k',\hat\lambda'}}^\dagger \right\rangle_1 \\
  \equiv  \sum_{\hat A',A'} D_{\hat A',A'}  (0)
   \langle 0 | {A'}^{a'}_{k',\lambda'} |A' \rangle 
   \langle \hat{A}' | {{A'}^{\hat a'}_{\hat k',\hat\lambda'}}^\dagger | 0 \rangle \\
  = \sum_{\hat A',A'} D_{\hat A',A'}  (0)
   \mathcal{A'}^{a'}_{\lambda'}(k',\lambda') 
   {\mathcal{A'}^{\hat a'}}^*(\hat k',\hat\lambda') \\
  = \delta_{\hat\lambda' \lambda'}
   \delta_{\hat{a}' a'} (2\pi)^4 \delta^4(\hat k' - k') \mathcal{N} \zeta^2 
   e^{-\zeta^2({k'^0}^2+\mathbf{k'}^2)} .
\label{eq:Dnuc}
\end{multline}

In order to determine the normalization constant $\mathcal{N}$ we 
could calculate the energy density of gluons in the nucleus, 
$\rho_g \equiv \frac{E_g}{V} = \frac{1}{VT} \int dt \tr [HD]$, but we
will see later that our final result does not depend on $\mathcal N$.

Returning to the expression (\ref{eq:dfinal}) for the final-state density
matrix, we obtain
\begin{multline}
D_{\hat{A},A}(t) = g^2 N_c  
  \int\frac{d^4k'}{(2\pi)^4}
  {\mathcal P} \sum_{\lambda'} |\epsilon^- (k',\lambda')|^2 \\
  \times (2\pi)^2 \delta(k^- -k'^-) \delta (\hat k^- - k'^-) 
  \frac{\delta^2 (\mathbf{k}_\perp - \mathbf{\hat k}_\perp )}
          {2V\sqrt{k^+ \hat k^+}} \\
  \times 
\delta_{\hat{a} a}
G(|\mathbf{k}_\perp - \mathbf{k}'_\perp|)\,  
  {\mathcal N} \zeta^2 e^{-\zeta^2({k'^0}^2+\mathbf{k'}^2)} \, .
\end{multline}
The $\delta$-function enforces $k'^- = k^-$. 
Furthermore, we can use $k \approx \hat k$ to argue that 
$\mathcal{P} \approx \frac{1}{2} \delta_{\lambda \hat\lambda}$ because the 
projections in $\mathcal{P}$ are maximal if $\lambda = \hat\lambda$ and 
the average value upon integration over the directions of 
$\mathbf{k'}_\perp$ should be $\langle \cos^2 \phi \rangle \approx 1/2$. 
The same argument also allows us to use the approximation 
$\sum_{\lambda'} |\epsilon^- (k',\lambda')|^2 \approx 3/4$,
allowing for three polarization states of the off-shell gluons in nucleus 1.
We can thus write
\begin{align}
D_{\hat{A},A} =  & \frac{3g^2N_c \zeta{\mathcal N}}{32\pi^{3/2}} 
  \delta_{\hat{a} a} \delta_{\hat\lambda \lambda} 
  e^{-\zeta^2{(k^-)}^2}
  \nonumber \\
  & \times \delta(k^- - \hat k^-) 
  \frac{\delta^2(\mathbf{k}_\perp - \mathbf{\hat k}_\perp)}
          {2V\sqrt{k^+ \hat k^+}} 
  \nonumber \\
  & \int d^2k'_\perp G(|\mathbf{k}_\perp - \mathbf{k'}_\perp|)  
  e^{-\zeta^2{k'_\perp}^2} .
\end{align}
We now introduce the convolution of the gluon two-point function with
the Gaussian profile 
\begin{equation}
  F({k}_\perp) = \int d^2 k'_\perp G(|\mathbf{k}_\perp - \mathbf{k'}_\perp|)
  e^{-\zeta^2{k'_\perp}^2} ,
\end{equation}
and thus obtain our final expression for the final-state gluon density
matrix (\ref{eq:dfinal}):
\begin{equation}
  D_{\hat A,A} =  {\mathcal N} \frac{3\alpha_s N_c\zeta}{8\sqrt{\pi}V}
    \delta_{\hat\lambda \lambda}\delta_{\hat{a} a} 
    \frac{\delta^3(k-\hat k)}{2\sqrt{k^+ \hat k^+}} 
    e^{-\zeta^2{(k^-)}^2} F({k}_\perp) ,
\end{equation}
where the three-dimensional delta function refers to the ``$-$'' and
``$\perp$'' components of the momenta. We note that the density matrix 
resulting from the interaction with the external gluon field for large
times is diagonal in all quantum numbers except for the longitudinal momentum, 
in accordance with the result obtained in ref.~\cite{Muller:2005yu}.

Now we proceed to calculate the traces.
\begin{multline}
  \tr D  = \sum_A D_{A,A} = VT \int\frac{d^4k}{(2\pi)^4} 
  \sum_{a,\lambda} D_{A,A}  \\  
= T \delta^3(0) \mathcal{N}
  \frac{3\alpha_s N_c(N_c^2-1)}{32\pi^2} 
  \int\frac{dk^+}{k^+} \int \frac{d^2 k_\perp}{(2\pi)^2} 
  F({k}_\perp) \, .
\end{multline}
The integral over $k^+$ should be regulated by a phase space projection,
because the gluons are (almost) on the mass-shell, but there is no need to 
specify the details here. We have also assumed that the final-state gluons can 
carry only transverse polarizations, because they are nearly on mass-shell. 
The trace of the square is
\begin{multline}
  \tr D^2 = \sum_{A,A'} D_{A,A'} D_{A',A} \\
  = T^2\delta^3(0) \mathcal{N}^2 
     \frac{9\alpha_s^2 N_c^2 (N_c^2-1)\zeta}{128\sqrt{2\pi}(2\pi)^6} 
     \\ \times
     \left(\int\frac{dk^+}{k^+}\right)^2 \int \frac{d^2 k_\perp}{(2\pi)^2} 
     F({k}_\perp)^2
\end{multline}
For the ratio that we are seeking this leads to the expression.
\begin{equation}
  \frac{\tr D^2}{[\tr D]^2} = \frac{1}{(2\pi)^3\delta^3(0)} 
  \frac{\zeta\sqrt{2\pi}}{2(N_c^2-1)}
  \frac{I_2}{(I_1)^2}
\label{eq:D2D}
\end{equation}
where 
\begin{align}
 I_1 & = \int d^2 k_\perp g^2 F({k}_\perp) \\
 I_2 & = \int d^2 k_\perp g^4 F^2({k}_\perp)
\end{align}
We note that the volume associated with 
$\delta^3(0) = \delta(0^-)\delta^2(\mathbf{0}_\perp)$
is proportional to the light-cone ``time'' variable $x^+$, which is conjugate
to $k^-$. When boosted to mid-rapidity, $x^+$ transforms as
$x'^+ = x^+/(2\gamma)$, unlike the regular time coordinate $x^0$, 
which transforms as $x'^0 = \gamma(x^0-\beta x^3)$.
This fact is in line with our intuitive picture in analogy to Coulomb-explosion:
Decoherence occurs due to the fact that Lorentz contracted 
nucleus 2, which acts as if consisting of incoherent color-charges, 
passes through nucleus 1.   

The result we obtained coincides with that of our previous calculation 
in predicting 
a decoherence behavior $\sim 1/x^+$ at leading order \cite{Muller:2005yu}. 
It is given as a function of $F(\mathbf{k}_\perp)$ which can be evaluated 
using different models for the initial two-gluon correlator $G$.

\section{Decoherence in the McLerran-Venugopalan Model}

In this section we want to compute $I_1$ and $I_2$ using the standard 
two-point gluon function from the McLerran-Venugopalan model. 
We work with the Fourier transform
$G(\mathbf{p})  =  \int d^2 x \, e^{-i\mathbf{px}} f(\mathbf{x})$. Here and 
in the following we suppress the index ``$\perp$'' for easier notation.
$f$ has first been calculated in the McLerran-Venugopalan model
in \cite{JKMW:96}. We follow the conventions in Lappi \cite{Lappi:06}
and write 
\begin{multline}
  f(\mathbf{x}) = \frac{4 (N_c^2-1)}{N_c g^2 x^2} \\ \times
  \left( 1- e^{ -g^4 N_c/(8\pi)\mu^2 x^2 \ln 1/( x\Lambda)} \right)
\end{multline}
where $\mu^2$ is related to the saturation scale $Q_s \sim g^2\mu$.
This result is only valid for $x<1/\Lambda$ where $\Lambda$ is a IR cutoff
and we set $f=0$ for $x>1/\Lambda$. All vectors are 2-vectors in the
transverse plane.

Note that we have defined factors of the coupling constant $g$ into
$I_1$ and $I_2$ without canceling them in the ratio. We do so because
only the square of the gluon field strength tensor times the running
coupling $\alpha_s$ has well-defined properties and we consider 
$g^2 G(p)$ to be the physical quantity. The correct implementation
of the running coupling is a topic 
of intense investigations.
We follow the prescription by Kovchegov and Weigert \cite{Kovchegov:2007vf}
and substitute
\begin{equation}
g^4 ~\to g^2(\Lambda^2) g^2(1/x^2)
\end{equation} 

For $I_1$ we obtain
\begin{multline}
  I_1 = (2\pi)^2 \frac{\pi}{\lambda^2} \lim_{x\to 0} (g^2 f(\mathbf{x}))
  \\ =
  (2\pi)^2 \frac{\pi}{\lambda^2} \lim_{x\to 0} 
  \frac{N_c^2-1}{2\pi} \mu^2 g^2(\Lambda^2) g^2(1/x^2) \ln 1/(x\Lambda) \\
  = (2\pi)^4 g^2(\Lambda^2) \frac{(N_c^2-1) \mu^2}{\beta_0 \lambda^2} 
\end{multline}
where we used the 1-loop running coupling with
\begin{equation}
  \beta_0~=~ \frac{11}{3} N_c -\frac{2}{3} N_f \, .
\end{equation} 
Obviously this is a well defined expression, while $I_1$ without the
running coupling would have led to a logarithmic UV divergence.

On the other hand, after two Gaussian integrations we see that
\begin{multline}
  \label{eq:i2}
  I_2 = (2\pi)^4 \frac{\pi^2}{\lambda^4} \int \frac{d^2 x}{(2\pi)^2} 
  e^{-x^2/(2\lambda^2)} g^4 f^2(x) \\
  = (2\pi)^3 \frac{\pi^2}{\lambda^4} \frac{16(N_c^2-1)^2}{N_c^2}
   \int_0^{\Lambda^{-1}} \frac{dx}{x^3} e^{-x^2/(2\lambda^2)} \\
  \times \left[1- \exp\left( -g^2(\Lambda^2)\mu^2 x^2 
   \frac{\pi N_c \ln (1/x^2\Lambda^2)}{\beta_0 
   \ln(1/x^2\Lambda_{\mathrm{QCD}}^2)}    \right)\right]^2
\end{multline} 
The scale $\Lambda$ should be chosen such that for typical transverse
momenta $1/x$ ideally fulfills \cite{Kovchegov:2007vf}
\begin{equation}
\frac{1}{x} \gg \Lambda \gg \Lambda_{QCD} \, .
\end{equation}
For $x$ between 0 and $1/\Lambda$ the ratio of the logarithms should
hence obey
\begin{equation}
0\leq \ln(1/x^2\Lambda^2)/\ln(1/x^2\Lambda_{QCD}^2) \leq 1 ,, .
\end{equation}
We conclude that $I_2 < \tilde I_2$ where $\tilde I_2$ is given by the
last expression in Eq.\ (\ref{eq:i2}) with the ratio of logarithms 
replaced by one. This inequality is useful because the remaining integral 
in $\tilde I_2$ can be solved analytically.

After replacing $u=x^2\Lambda^2$ and introducing the short notations
$a=(2\lambda^2\Lambda^2)^{-1}$ and $b=\pi N_c\mu^2 g^2(\Lambda^2)/
(\Lambda^2 \beta_0)$ the integral is
\begin{multline}
  \tilde I_2 = 2(2\pi)^5 \frac{\Lambda^2}{\lambda^4} \frac{(N_c^2-1)^2}{N_c^2}
  \int_0^1 \frac{du}{u^2}  \\ \times
  \left( e^{-au}-2e^{-(a+b)u}+e^{-(a+2b)u} \right)\, .
\end{multline}
After two partial integrations this can be brought into the form
\begin{multline}
  \tilde I_2 =  2(2\pi)^5 \frac{\Lambda^2}{\lambda^4} \frac{(N_c^2-1)^2}{N_c^2}
  \\ \times \left[ -\left(e^{-au} -2e^{(a+b)u} + e^{-(a+2b)u}
    \right)   \right.
  \\ - \left. \int_0^1 du \ln u \left( a^2 e^{-au} 
   - 2(a+b)^2 e^{(a+b)u}\right.\right.  \\  + \left. \left. 
     (a+2b)^2 e^{-(a+2b)u} \right) \right].
\end{multline}
Now we expand the integration region of the remaining integral to infinity,
anticipating that the integrand is rapidly vanishing for $u\to \infty$.
Then the integral gives
\begin{multline}
  \tilde I_2 =  2(2\pi)^5 \frac{\Lambda^2}{\lambda^4} \frac{(N_c^2-1)^2}{N_c^2}
  \\ \times \left[ -e^{-au} +2e^{(a+b)u} - e^{-(a+2b)u}  \right. \\ 
  +  a \left( \gamma+\ln a \right) - 2(a+b)\left( \gamma + \ln (a+b)\right) 
  \\ + \left. (a+2b)\left( \gamma +\ln (a+2b) \right) \right]
\end{multline}
where $\gamma$ is Euler's constant.

In the region of applicability for the Color Glass Condensate we 
expect $b \gg a \sim \mathcal{O}(1)$. Hence the bracket in the last
equation is to good approximation equal to $2b \ln 2$.
Finally we obtain 
\begin{equation}
  \tilde I_2 \approx 2\ln 2 (2\pi)^6 g^2(\Lambda^2) 
  \frac{\mu^2}{\beta_0 \zeta^4} \frac{(N_c^2-1)^2}{N_c} \, ,
\end{equation}
giving the following bound for the relevant ratio of integrals:
\begin{equation}
  \frac{I_2}{(I_1)^2} < \frac{2\beta_0\ln 2}{(2\pi)^2 g^2(\Lambda^2) N_c \mu^2} \, .
\end{equation}

We now return to Eq.~(\ref{eq:D2D}).
The expression $(2\pi)^3\delta^3(0)$ in the denominator 
gives the transverse  normalization area times the observation time $T$. 
To fix the transverse normalization area one can follow two different
lines of argument. 
The incoming gluons from nucleus 1 are effectively localized within 
the transverse area $\pi \zeta^2$ given by the initial density matrix 
(\ref{eq:Dnuc}). If the density matrix $D$ is interpreted as 
that of a completely coherent system $\tr D^2 \approx (\tr D)^2$ than 
the transverse normalization area has to be chosen as $\pi \zeta^2$.
If, on  the other hand, one prefers to extend the normalization area 
to the whole area of the nucleus,$\pi R^2$, one has to take into account 
that the starting value of  
$\tr D^2 / (\tr D)^2$ is not close to one but rather of the order 
$\zeta^2/R^2$ and one should thus ask after which time 
$R^2\tr D^2 / \zeta^2 (\tr D)^2$ has dropped to $1/e$. 
(In the latter case $D$ has the form of a block-diagonal matrix with
$R^2/\zeta^2$ blocks.)
It is reassuring, that both 
lines of argument lead to the same result.
\begin{equation}
\frac{R^2}{\zeta^2} \frac{\tr D^2}{[\tr D]^2} 
 < \frac{2\beta_0\ln 2}{(2\pi)^{5/2} N_c (N_c^2-1) 
  g^2(\Lambda^2) \mu^2 } \frac{1}{\zeta T}
\end{equation}
Defining the decoherence time $\taudec$ as the time where this ratio 
has dropped to a value $1/e$ and fixing the physical saturation scale as 
$Q_s = g^2(\mu^2) \mu$, we obtain the upper bound
\begin{eqnarray}
\taudec 
&<& \left(\frac{8 e \ln 2}{\sqrt{2\pi}N_c(N_c^2-1)}\right)
         \left(\frac{g^2(\mu^2)}{g^2(\Lambda^2)}\right) 
         \nonumber \\
  & & \left(\frac{\beta_0 g^2(\mu^2)}{16\pi^2}\right)
         \left(\frac{1}{\zeta Q_s}\right) \frac{1}{Q_s} 
         \nonumber \\
&\approx & 0.25  \left(\frac{g_\mu^2}{g_\Lambda^2}\right) 
         \left(\frac{\beta_0 g_\mu^2}{16\pi^2}\right)
         \left(\frac{1}{\zeta Q_s}\right) \frac{1}{Q_s} \, .
\end{eqnarray}
All factors in parentheses being of order unity, we thus conclude
that $\taudec \sim Q_s^{-1}$ in agreement with the result obtained
in Ref.~\cite{Muller:2005yu}.

\section{Decoherence Entropy}

We now turn to the question how much entropy can be produced
by the rapid 
decoherence of the initially coherent nuclear gluon field.
In order to illustrate the mechanism, we first discuss a simple model 
for which the relevant calculations can be performed exactly \cite{deco3}, 
but which is sufficiently general to permit a semi-quantitative estimate 
of the entropy produced by decoherence in a heavy ion reaction.

The quantum mechanical analogue of a classical field is a coherent
state \cite{Glauber}
\begin{equation}
| \Psi[J] \rangle = \prod_{\bf k,\lambda} \exp( i\alpha_{{\bf k}\lambda} 
          a_{{\bf k}\lambda}^\dagger  - i\alpha_{{\bf k}\lambda}^* 
          a_{{\bf k}\lambda} ) | 0 \rangle ,
\label{eq1}
\end{equation}
where the amplitude $\alpha_{{\bf k}\lambda}$ is determined by the classical
current ${\bf J}$ creating the field
\begin{equation}
\alpha_{{\bf k}\lambda} = (\hbar\omega_{\bf k}V)^{-1/2} 
          {\bf \epsilon}_{{\bf k}\lambda} \cdot 
          {\bf J}({\bf k},\omega_{\bf k}) .
\label{eq2}
\end{equation}

Let us begin by considering a single mode ${\bf k}\lambda$. The coherent
state can be written as a superposition of particle number eigenstates:
\begin{equation}
|\alpha\rangle = e^{-|\alpha|^2/2} \sum_{n=0}^\infty 
          \frac{\alpha^n}{\sqrt{n!}} |n\rangle .
\label{eq3}
\end{equation}
Being a pure quantum state, $|\alpha\rangle$ is described by a density
matrix
\begin{equation}
\rho_{mn} = \langle m|\alpha\rangle \langle\alpha |n\rangle ,
\label{eq4}
\end{equation}
which satisfies the relation $\rho^2 = \rho$ and has no entropy:
$S = - {\rm Tr}\,\rho\, \ln\rho = 0$. 

Complete decoherence of this quantum state corresponds to the total 
decay of all off-diagonal matrix elements of the density matrix, 
yielding the diagonal density matrix
\begin{equation}
\rho^{\rm dec}_{mn} = |\langle n|\alpha\rangle|^2 \delta_{mn} =
          e^{-|\alpha|^2} \frac{|\alpha|^{2n}}{n!} \delta_{mn} .
\label{eq5}
\end{equation}
The particle number in this mixed state follows the Poisson distribution,
and the average number of particles is ${\bar n} = |\alpha|^2$. 
The entropy content of the mixed state is given by
\begin{eqnarray}
S_{\rm dec}^{\rm (cs)} & = & \sum_{n=0}^{\infty} 
          e^{-\bar n}\frac{{\bar n}^n}{n!}
          \ln\left(e^{-\bar n}\frac{{\bar n}^n}{n!}\right)
          \nonumber \\
          & = & e^{-{\bar n}} \sum_{n=0}^\infty \frac{{\bar n}^n}{n!} 
          (n \ln {\bar n} - {\bar n} - \ln n!) ,
\label{eq6}
\end{eqnarray}
where the superscript ``cs'' indicates that the result holds for a
coherent state. With the help of Stirling's formula and the integral 
representation of the logarithm,
\begin{equation}
\ln n = \int_0^{\infty} \frac{ds}{s} \left( e^{-s} - e^{-ns} \right) ,
\label{eq7}
\end{equation}
the sum in (\ref{eq6}) can be performed yielding an analytical result 
that is valid asymptotically for ${\bar n} \gg 1$ (actually, the
approximation is excellent already for ${\bar n} \approx 1$):
\begin{equation}
S_{\rm dec}^{\rm (cs)} = \frac{1}{2} \left( \ln (2\pi {\bar n}) \, + 1 -
          \frac{1}{6{\bar n}} + \cdots \right) .
\label{eq8}
\end{equation}
It is not surprising that the entropy is proportional to 
$\ln \sqrt{\bar n}$, because we have deleted all information about 
the relative signs of the amplitudes $\langle\alpha |n\rangle$ by 
eliminating the off-diagonal elements of the density matrix. 
The number of significantly contributing elements is given by the
width, $\Delta n = \sqrt{\bar n}$, of the Poisson distribution.
That the decoherence entropy is controlled by $\Delta n$, rather
than by $\bar n$, can be seen by considering more general pure 
quantum states, for which the average occupation number $\bar n$ and 
the occupation number uncertainty $\Delta n$ are not related. For a 
pure state with ${\bar n} \gg \Delta n \gg 1$ and in the Gaussian 
approximation, it is straightforward to show that the decoherence 
entropy is given by
\begin{equation}
S_{\rm dec} = \frac{1}{2} \left( \ln (2\pi(\Delta n)^2) \, 
          + 1 + \cdots \right) ,
\label{eq8a}
\end{equation}
confirming our assertion. For a classical coherent state (\ref{eq3}), 
the expression (\ref{eq8a}) coincides with (\ref{eq8}).

\begin{figure}[tb]   
\resizebox{0.95\linewidth}{!}
          {\rotatebox{-90}{\includegraphics{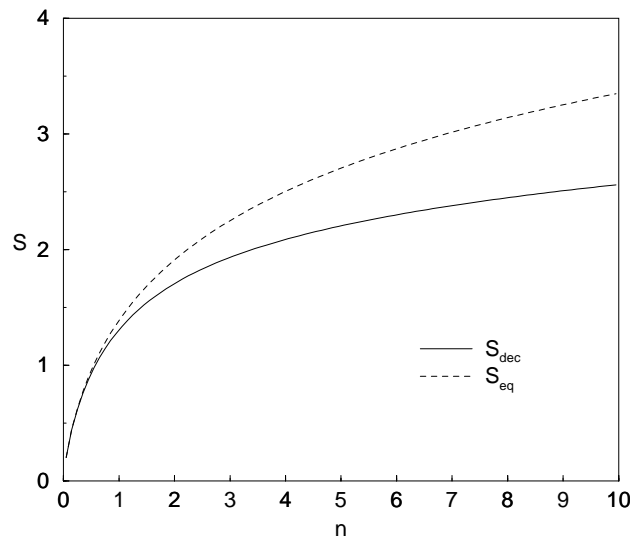}}}
\caption{Decoherence entropy $S_{\rm dec}$ for a coherent state of a 
         single field mode and equilibrium entropy $S_{\rm eq}$ for 
         the same average total energy as a function of the average 
         occupation number $\bar n$.}
\label{fig1}
\end{figure}

We also note that the entropy for a single quantum oscillator 
in equilibrium at temperature $T$ is given by
\begin{equation}
S_{\rm eq} = \ln ({\bar n}+1) + 
             {\bar n}\ln\left(1+\frac{1}{\bar n}\right) ,
\label{eq9}
\end{equation}
where ${\bar n} = (e^{\omega/T}-1)^{-1}$ is the average occupation
number. Asymptotically, for large $\bar n$, one obtains $S_{\rm eq}
\approx 2 S_{\rm dec}^{\rm (cs)}$, i.\ e.\ the thermal entropy becomes
twice as large as the decoherence entropy. However, for small to
moderate occupation numbers the ratio $S_{\rm dec}^{\rm (cs)}/S_{\rm eq}$ 
is close to unity. Figure \ref{fig1} shows the decoherence and equilibrium 
entropies as a function of the average occupation number $\bar n$. 
For not too large values of $\bar n$, the decoherence process generates 
a large fraction of the equilibrium entropy,
and any subsequent equilibration process adds only a small amount of
entropy to it. 

What does this imply for quantum field theory, where the field is
a system of infinitely many coupled oscillators? Assume that, after 
decoherence, the system can be described as a collection of $N$
particles, given by some distribution function over single-particle
states, which were generated by the decoherence of $N_{\rm cs}$ 
coherent quantum states. Examples of such states are the internal
wavefunctions of nucleons forming a large nucleus, or a quark with
its comoving gluon cloud. Each coherent state contributes on average
${\bar n} = N/N_{\rm cs}$ partons. Then, after full equilibration,
the thermal entropy is of the order of $S_{\rm th} \sim N_{\rm cs}
{\bar n} = N$, while for the decoherence entropy we get
$S_{\rm dec} \sim N_{\rm cs} \frac{1}{2} \ln(2\pi{\bar n})$.
The ratio of the two entropies is
\begin{equation}
\frac{S_{\rm dec}}{S_{\rm th}} \sim \frac{\ln(2\pi{\bar n})}{2 \bar n} ,
\label{eq10}
\end{equation}
i.\ e.\ for large amplitude quantum states, which turn into many
particles per coherent mode, the decoherence contribution to the
thermal entropy is small. On the other hand, if the individual
occupation numbers are of order one, the contribution is sizable.
This case applies to our problem of interest, the collision of two 
nuclei at high energy, as we will discuss in the following.

For the coherent color fields in colliding nuclei, the average number
of decohering gluons per transverse area has been given by \cite{Kov}
\begin{equation}
\frac{dN}{d^2bdy} \approx
           \frac{C_{\rm F}\ln 2\, Q_s^2}{\pi^2\alpha_s} .
\label{eq11}
\end{equation}
where 
$C_{\rm F}=4/3$. 
The characteristic transverse 
area, over which the color fields in nucleus 2 are coherent, 
is $\pi/Q_s^2$, and one can
argue that the longitudinal coherence length is of the order of
$\Delta y \approx 1/\alpha_s$ \cite{KLM01}. We thus obtain for the average
number of decohering partons per coherence domain
\begin{equation}
{\bar n} = \frac{dN}{d^2bdy} \frac{\pi}{Q_s^2} \Delta y
\approx \frac{C_{\rm F}\ln 2}{\pi\alpha_s^2} \approx 3 .
\label{eq12}
\end{equation}
For this value, our arguments presented above indicate that the entropy 
produced in the decoherence process is about half of the equilibrium
entropy. Applying Eq.~(\ref{eq8}) and using that the initial number of 
coherent domains per transverse area is $(Q_sR)^2$, we find that the 
total entropy per unit rapidity produced by decoherence in a Au+Au 
collision at RHIC is
\begin{eqnarray}
\frac{dS_{\rm dec}}{dy} & \approx & \frac{Q_s^2 R^2}{2 \Delta y} 
          \left(\ln(2\pi {\bar n}) +1\right)
          \nonumber \\
          & \approx & \frac{Q_s^2 R^2 \alpha_s}{2} \left[
          \ln \frac{2C_{\rm F}\ln 2}{\alpha_s^2} + 1 \right]
          \approx 1500 ,
\label{eq13}
\end{eqnarray}
where we used the values \cite{Kov} $Q_s^2\approx 2$ GeV$^2$, $R=7$ fm,
and $\alpha_s \approx 0.3$. This value accounts for about one-third of the
entropy measured in the final hadron distribution \cite{Pal:2003rz,Muller:2005en}.

\section{Conclusions}

We advocate the idea that a large fraction of the total entropy produced
in high-energy heavy-ion collisions is generated by decoherence
of the many-body quark-gluon wave functions of the colliding nuclei 
in the very first phase of the collision. 

We presented an improved determination of the decoherence time 
$\taudec$ as a function of the initial gluon correlation function. 
Within the color-glass-condensate formalism this leads to a
decoherence time $\taudec \leq 1$ fm/$c$ which agrees with the result 
of an earlier calculation. We also estimate the entropy produced through
decoherence of the initial gluon field and find that it could contribute
about one third of the total entropy observed at RHIC.

\section{Acknowledgments}
This work was supported by the Alexander von Humboldt Foundation, 
BMBF, RIKEN/BNL, the Texas A\&M College of Science, and DOE grant 
DE-AC02-98CH10886.

\end{document}